\documentstyle[twocolumn,floats,aps,epsf,times]{revtex}

\begin{document}

\newcommand{\be}{\begin{eqnarray}}
\newcommand{\en}{\end{eqnarray}}
\newcommand{\mc}{\mathcal}
\newcommand{\no}{\nonumber}
\newcommand{\ie}{\mbox{\protect{\it i.\ e.\ }}}
\newcommand{\eg}{\mbox{\protect{\it e.\ g.\ }}}
\newcommand{\cf}{\mbox{\protect{\it c.\ f.\ }}}
\newcommand{\etal}{\mbox\protect{\it et.\ al.\ }}

\twocolumn[
\hsize\textwidth\columnwidth\hsize
\csname@twocolumnfalse\endcsname

\draft
\title{
XPS as a Probe of Gap Opening in Many Electron Systems
}
\author{Robert Haslinger and Nic Shannon}
\address{
Department of Physics, University of Wisconsin--Madison,
1150 Univ. Av., Madison WI 53706--1390, USA.
}
\date{\today}

\maketitle

\begin{abstract}
Core hole photoemission (XPS) provides a powerful indirect
probe of the low energy excitations of a many electron
system.   We argue that XPS can be used to study the way
in which a gap opens at a metal--superconductor or
metal--insulator transition.   We consider the ``universal''
physics of how the loss of low energy excitations modifies XPS
spectra in the context of several simple models, considering
in particular the case of a two dimensional d--wave
superconductor.
\end{abstract}


\vspace*{\baselineskip}

]
\narrowtext

\section{Introduction}
\label{intro}

As a many electron system undergoes a phase transition, the nature of
its low energy excitations is usually radically altered.
For this reason, experiments which are sensitive to the
rearrangement, and in particular to the loss of,
low energy excitations (to the opening of a gap), can
inform us about how phase transitions take place.
Nuclear Magnetic Resonance (NMR) has been used widely in this context to
study how low lying {\it spin} excitations evolve in different phases.
Famously, NMR reveals a second ``transition'' temperature $T^*$ in the
underdoped cuprate superconductors, below which a gap opens to spin
excitations, even in the absence of superconducting order
\cite{slichter}.

In several previous articles \cite{taisyii} \cite{nic} one of us proposed the use of
core level photoemission (XPS) as a complimentary probe to NMR, offering
similar insight into the evolution of {\it charge} carrying
excitations.  In the case of the underdoped cuprates, the comparison
of XPS and NMR might shed some light on the existence or absence of
{\it spin--charge separation} by answering the question of whether
or not a gap opens to charge excitations at the spingap temperature $T^*$.
In fact there are many senarios in which such a comparison might
be informative --- for example in metal--insulator transitions
where electron--electron interaction is strong.

The opening of a charge gap in a metallic system has certain simple
systematic consequences for XPS lineshapes --- an overall shift in the
core line, and a transfer of spectral weight from high energies to the
line threshold.
These effects, both due to the suppression of low energy density
fluctuations can be understood on quite general physical grounds
without recourse to specific models or calculations \cite{OCrevisited}.

In this article, we develop a more quantitative picture
of how XPS lineshapes evolve with the opening of a gap
by applying perturbation theory to simple
models of a metal--semiconductor and metal--superconductor
transitions.   A good understanding of the behaviour of the simplest
models is clearly a necessary first step for the interpretation of
experiment in more complicated material examples, such as the
cuprates.
With this in mind we make a detailed case study of the
of the way in which the familiar asymmetric Doniach--Sunjic lineshape
for a core level in a metal
\cite{doniach} is modified by the opening of a
superconducting gap in s-- and d--wave superconductors,
discussing the role of nodes in the gap.

We begin in Section \ref{formalism} with an outline of the simple perturbative
formalism used to calculate lineshapes and associated shifts.
In Section \ref{band} we calculate the XPS spectrum for a band metal
and a simple toy model of a semiconductor within perturbation theory.
In Section \ref{sc} we consider lineshapes and shifts
of s-- and d--wave superconductors for a free electron gas and
a d--wave superconductor on a two dimensional tight binding lattice at half
filling.

The qualitative picture for the evolution of the XPS lineshape
set out in \cite{OCrevisited} is found to hold; differences in detail
for different models are explored quantitatively.
In the concluding Section \ref{conclusions} we discuss the
consequences and limitations of these results, as applied to
experiment, emphasizing the potential role of XPS as
a diagnostic tool for strongly correlated systems.

\section{Formalism}
\label{formalism}

In an XPS experiment a high energy (X--Ray) photon ejects
a single electron from a tightly bound atomic level in
the sample material, typically a prepared metal surface.
The energy distribution of the emited photoelectrons is measured.
Within the sudden approximation, and neglecting all momentum
dependence of the matrix elements, the XPS lineshape for the
core level is simply proportional
to the spectral function for the resulting core hole
\cite{photoemission}.
Because photoemission leaves behind this unscreened and massive
positive charge (recoil of the core hole can safely be neglected)
it is accompanied by a violent low energy "shake up" of the remaining
itinerent electrons.   This manybody effect has important
consequences for the core lineshape, as described below, and it
is the suppression of the ``shake up'' by a gap to charge
excitations which makes XPS useful as a probe of different phases.

In practice various other mechanisms serve to limit the lifetime
of the core hole --- which must eventually be filled by the
decay of an electron from a higher energy  level --- and we model
these by convoluting the calculated lineshape with a Lorentzian
of width the inverse core level lifetime.  We neglect a further
(temperature dependent) broadening due to phonon processes, which
could be accounted for by further convolution with a Gaussian.
Perhaps more importantly, we do not consider the case on which the
core hole binds an itinerent electron.  This leads to the possibility
of more than one line accompanying each core level, which has been
treated for the ordinary free electron gas by several authors
\cite{combescot}.

We model the combined system of core level and
itinerent electrons with the simple Hamiltonian
\begin{equation}
{\mc H} = {\mc H}_0 + {\mc V}_c
\end{equation}
where ${\mc H}_0$ is the unperturbed Hamiltonian describing
the itinerent electron system, and
\begin{eqnarray}
{\mc V}_c &=& \epsilon_d d^{\dagger} d + V(t) \nonumber\\
&=& \epsilon_d d^{\dagger} d + \frac{1}{\nu^2}\sum_{k,q}
V(q) c^{\dagger}_{k-q}(t) c_k(t)
\end{eqnarray}
switched on suddenly at t=0 when the core hole is created.
$\epsilon_d$ is the energy of the core hole,
$d$ is the core hole annihilation operator and $c$ the electron
annihilation operator.  Spin does not enter into the problem
and has been supressed in our notation.

We model gapless systems as a band of spinless
non--interacting electrons
\be
{\mc H}_0 &=&
   \sum_{k} \epsilon_k c^{\dagger}_{k}c_{k}
\en
characterized by a density of states
$N(\omega) = \sum_k \delta(\omega - \epsilon_k)$.
Semi--conductors are modeled in the same way, but
with zero density of states within the gap $|\omega|<\Delta$.
We use the usual BCS description of superconducting systems,
with quasi--particle dispersion
$E_k^2 = \epsilon_k^2 + \Delta_k^2$, where $\epsilon_k$ is
the underlying band dispersion and $\Delta_k$ the (momentum
dependant) superconducting order parameter.

Within the sudden approximation \cite{photoemission}, the XPS
spectrum is proportional to the core electron spectral function
\be
A_h(\omega) = -2 Im \left\{ G_h^{ret}(\omega)  \right\}
\en
where $G_h^{ret}(\omega)$ is the
retarded core hole Green's function.  From this definition
it follows that the spectral function is normalized to $2\pi$, and
spectral weight must always be conserved in XPS lineshapes.

The Green's function for the core hole  must be calculated
using the full wavefunction for the many electron system
{\it including} the itinerent electrons, and therefore
involves matrix elements for the overlap of the many--electron
groundstate with all the different states excited by the suddenly switched
core hole.  In this indirect way XPS probes the spectrum of the
intinerent electron liquid.

In the absence
of any interaction with the core hole ($V(q) \equiv 0 $)
the itinerent electrons remain in their ground state and
the core hole spectral function is a single coherent
delta function peak
\be
A_h(\omega) = 2\pi \delta(\omega - \epsilon_d)
\en
Interaction with itinerent electrons transfers spectral
weight to an incoherent tail and,
under certain conditions,
eliminates the coherent (delta function) part of the spectral function
entirely.

We evaluate $G_h^{ret}(t)$ in the presence of interaction ($V(q) \ne 0 $)
using a linked cluster expansion \cite{mahan}
\be
G_h^{ret}(t) = -i\theta(t) e^{-i\epsilon_d t}
\exp\left[\sum_{l=1}^{\infty} F_l(t) \right]
\en
where the  coefficents $F_l(t)$ are given by:
\be
F_l(t)=\frac{1}{l}(-i)^l \int_0^t dt_1 ... \int_0^t dt_l
\langle \mid T V(t_1) ... V(t_l) \mid \rangle_{connected}
\en

The leading term in this series, $F_1(t)$, is purely real and
contributes only an absolute shift in the core line.
For a purely local interaction ($V(q) = V_0$), this is simply proportional
to the density of electrons, and therefore unchanged by the
opening of a gap.

It is the second order term, $F_2(t)$, which contains interesting many
body physics and, to second order in $V(q)$, determines the XPS lineshape.
This is seen to be related to the density--density correlation
function (charge susceptibility), and indeed it can be rewritten as:
\be
F_2(t) &=& \frac{1}{\nu} \sum_q |V(q)|^2 \left\{ \right.
\frac{i}{2} \chi_{\rho}^{\prime} (q,\omega=0) t \nonumber\\
&+& \frac{1}{\pi} \int_0^{\infty} d\omega
\chi_{\rho}^{\prime\prime}(q,\omega)
\frac{1- e^{-i\omega t}}{\omega^2} \left. \right\}
\en
where $\chi_{\rho}^{\prime}$ and $\chi_{\rho}^{\prime\prime}$
are the real and imaginary parts of the retarded density--density
correlation function in frequency space \cite{langreth}.

The first part of this expression is an energy shift.
It {\it is} sensitive to the opening of a gap and for
delta function interaction it is
proportional to the real part of the local charge susceptibility.
The second, more complicated term, determines the
line shape.  We write this as :
\be
\label{lineshape}
\overline{F}_2(t) =
-\int_0^\infty R(\omega) \frac{1-e^{-i\omega t}}{\omega^2}
\en
where:
\be
R(\omega) = - \frac{1}{\pi}\sum_q |V(q)|^2
   \chi^{\prime\prime}_{\rho}(q,\omega)
\en
is a spectral representation of the perturbation.
The core hole---itinerent electron interaction $V(q)$
is short ranged, and for the purposes of this article
may be approximated by the purely local interaction $V(q) = V_0$,
so
\be
R(\omega) = - \frac{1}{\pi} |V_0|^2
   \sum_q \chi^{\prime\prime}_{\rho}(q,\omega)
\en

To find the XPS lineshape within these approximations
is then a matter of calculating
the imaginary part of the local density-density correlation
function as a function of frequency.
In the next section we work several examples, rederiving the
familiar asymmetric (Doniach--Sunjic) lineshape for a core level
in a metal and showing how it is modified by the opening of a gap
in the excitation spectrum.

\section{Band metals and semi--conductors}
\label{band}

\begin{figure}[tb]
\begin{center}
\leavevmode
\epsfxsize \columnwidth
\epsffile{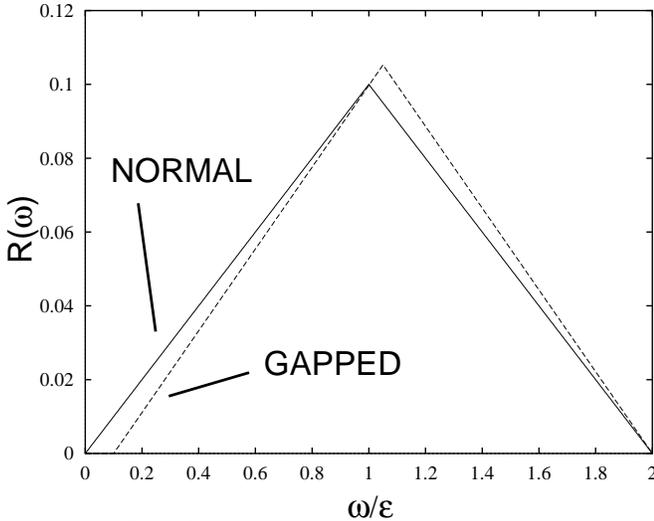}
\caption{
$R(\omega)$ for the constant density of states model,
with and without a gap of $\Delta/\epsilon=0.05$.  The
asymmetry exponent is $\alpha=0.1$.
}
\label{fig1}
\end{center}
\end{figure}

In this section we will calculate the lineshape for a normal metal,
modeled as a band of non--interacting
electrons %\cite{footnote}
using Eqn. \ref{lineshape}, and observe
how it changes when a gap is opened in the density of states.
For simplicity we assume zero temperature and a delta function
potential for the core hole.

Within a non--interacting picture, $R(\omega)$ can be found from the
imaginary part of the particle--hole bubble
\be
R(\omega)= \frac{|V_0|^2}{\nu^2} \sum_{k,p}
 [n(\xi_p) - n(\xi_k) ]\delta(\omega + \xi_p - \xi_k)
\en

We consider a flat density of states centered about the fermi
energy, such that :
\be
N(\omega) = \left\{
   \begin{array}{ll}
      N_0 & \quad \mid \omega \mid < \epsilon\\
      0   & \quad \mid \omega \mid > \epsilon\\
   \end{array} \right.
\en
$R(\omega)$ is then easily calculated.
\be
\label{metal}
R(\omega) = \left\{ \begin{array}{ll}
\alpha \omega & \mbox{$ 0<\omega<\epsilon$} \\
 \alpha (2\epsilon - \omega) & \mbox{$\epsilon<\omega<2\epsilon$} \\
 0 & \mbox{$\omega>2\epsilon$}
 \end{array}
 \right.
\en
where $\alpha= 2 N_0^2 |V_0|^2$ (the 2 is from the spin summation).
In what follows we will use $\alpha$ as a parameter, rather than
separately specifying the bare density of states $N_0$ and the
interaction strength $V_0$.   Since these are unchanged
by the opening of a gap, values of $\alpha$ found from experiments
on the metallic phase of materials can be used to parameterize
predictions for the their XPS lineshapes in a gapped phase.
Empirically, $\alpha \sim 0.1$ for most simple metals, and all
plots in this paper have been calculated for $\alpha = 0.1$.

\begin{figure}[tb]
\begin{center}
\leavevmode
\epsfxsize \columnwidth
\epsffile{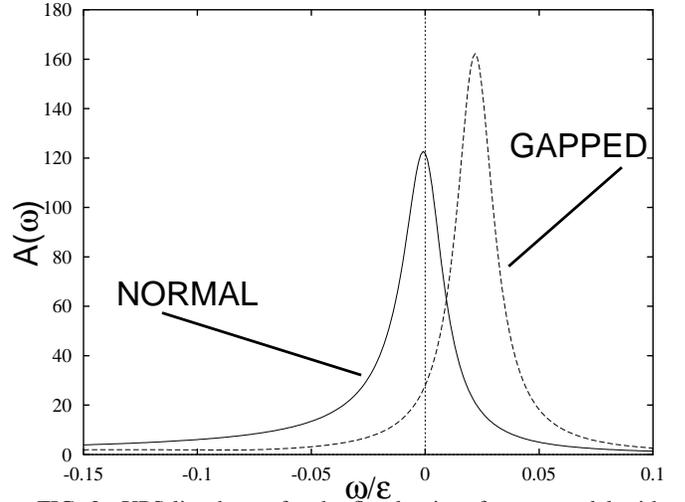}
\caption{XPS lineshapes for the flat density of states model
with and without a gap of $\Delta/\epsilon=0.05$.  The inverse core hole lifetime
is $1/\tau = 0.01\epsilon$, and $\alpha=0.1$.  The opening of a gap causes
spectral weight to be shifted out of the powerlaw tail and back
into the restored delta function peak.  In addition, the entire XPS
line shape (delta--function and incoherent tail) undergoes a rigid
shift to lower binding energy.
}
\label{fig2}
\end{center}
\end{figure}

On substitution of Eqn. \ref{metal} in Eqn. \ref{lineshape}
we find that the expected deltafunction at threshold is lost and
instead $A(\omega)$ diverges
as $\omega^{\alpha-1}$ as $\omega \to 0_+$.  This power law singularity
is broadened by the core hole lifetime, yielding an asymmetric lineshape
essentially equivalent to that calculated by Doniach and Sunjic
\cite{doniach}.  The XPS lineshape for this simple band metal
formed by numerically convoluting the spectral function with
a Lorentzian lifetime envelope is plotted in Fig. \ref{fig2}.
We have reversed the energy axis in this and all
other plots of lineshapes for ease of comparison with
photoemission spectra.

\begin{figure}[tb]
\begin{center}
\leavevmode
\epsfxsize \columnwidth
\epsffile{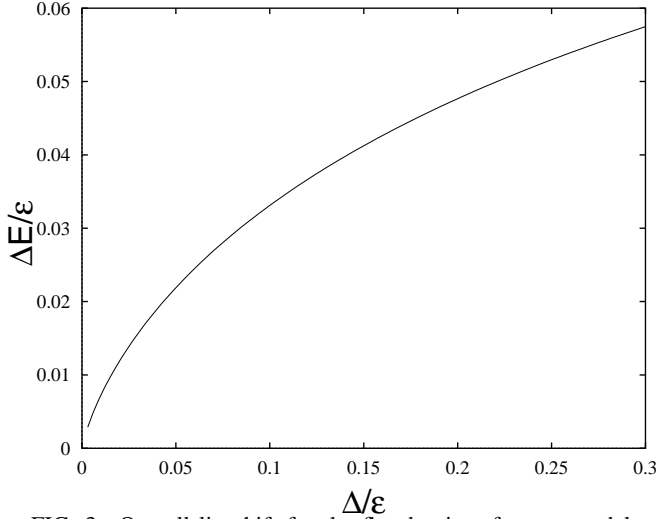}
\caption{Overall lineshift for the flat density of states
model at various gap magnitudes.  This is a shift away from the power
law tail, ie to lower binding energy.
($\alpha=0.1$).
}
\label{fig3}
\end{center}
\end{figure}

The replacement of the delta function peak in $A_h(\omega)$ with a power
law singularity is a consequence of Anderson's
{\it orthogonality catastrophy} \cite{anderson,hopfield}.
The sudden switching of the core hole in the photoemission process leads
to the creation of itinerent electron--hole pairs with all possible
energies and therefore to a high energy tail in the spectral function.
Since the number of electron--hole
pairs created with zero energy is logarithmically divergant,
the ground states of the perturbed and unperturbed systems are
orthogonal, and there is no delta function peak at threshold.
The orthogonality catastrophe is effective in a band of non--interacting
electrons whenever the density of states at the chemical potential
is finite, \ie for any band metal.
While all our analysis is limited to second order in the potential
${\mc V}$, this is usually small, and the physics of the orthogonality
catastrophe is in any case essentially
unaltered by the inclusion of higher order processes (multiple particle--hole
excitations) \cite{nozieres}.

We now calculate the lineshape for a toy model of a gapped system.
The presence of the gap will cut off the number of low energy
excitations made by the core hole, eliminating the orthogonality
catastrophe.  The gap has little effect at higher energies, so away
from threshold the lineshape for a system with a small gap
should be essentially unchanged.  The failure of the orthogonality
catastrophe will however lead to the restoration of a delta function
peak at threshold.  As the gap becomes bigger progressively more
spectral weight is transfered to this peak, so that for large
gaps the effective lineshape after convolution with a Lorentzian
will be the symmetric peak associated with an insulator.

At the same time the overall threshold for the XPS line is
shifted to lower energies because the redistribution of charge
in the electron gas is suppressed by the opening of the gap,
and therefore less work is done inserting a core hole into
the system.

If, for sake of illustration, we assume that a gap opens
such that the new density of states is :
\be
N(\omega) = \left\{\begin{array}{ll}
N_0 (\frac{\epsilon}{\epsilon-\Delta}) &
\mbox{$-\epsilon<\omega <-\Delta$} \\
N_0(\frac{\epsilon}{\epsilon-\Delta}) &
\mbox{$\Delta<\omega <\epsilon$}\\
0 & \mbox{otherwise}
\end{array}
\right.
\en
$R(\omega)$ is modified in a straightforward way
\be
R(\omega) = \left\{\begin{array}{ll}
0  & \mbox{$\omega < 2\Delta$} \\
\tilde\alpha(\omega - 2\Delta) &
\mbox{$2\Delta < \omega < \epsilon + \Delta$} \\
\tilde\alpha(2\epsilon - \omega)
&\mbox{$\epsilon + \Delta < \omega < 2\epsilon$}
\end{array}
\right.
\en
where $\tilde\alpha = (\epsilon/\epsilon-\Delta)^2\alpha$.
(Fig. \ref{fig1}).   This new form of $R(\omega)$ (imaginary part of the
charge susceptibility) completely determines the revised lineshape through
Eqn. \ref{lineshape}.

On the other hand it is the change in the real susceptibility which
leads to the shift in the XPS line.
The real and imaginary parts of the the density density correlation
function are connected via Kramers--Kronig relations.
\be
\chi_{\rho}^{\prime}(\omega)
   = \frac{1}{\pi}\int_{-\infty}^\infty \frac{
   d\tilde\omega \chi_{\rho}^{\prime\prime}(\tilde\omega)
   }{
   \tilde\omega - \omega}
\en
So the net line shift caused by the opening of the gap is
given by:
\be
\label{shift}
\Delta E &=& \int_0^\infty d\tilde\omega
   \frac{R_N(\tilde\omega) - R_G(\tilde\omega)}{\tilde\omega}
\en
This is easily calculated for our model system.
\be
\Delta E_{model} &=&  \alpha \lbrace
2\epsilon ~log2 - (\frac{\epsilon}{\epsilon-\Delta})^2
[ 2\epsilon ~log (\frac{2\epsilon}{\epsilon + \Delta})\no\\
   &+&
   2\Delta ~log \left(
   \frac{2\Delta}{\epsilon + \Delta}
   \right) ]\rbrace
\en
We plot this shift as a function of gap in Fig. \ref{fig3}
It is a shift {\it away} from the
power law tail, \ie towards lower binding energy.

A plot of the modified spectral function, convoluted
with a Lorentzian to mimic the finite core lifetime,
is shown in Fig \ref{fig2}.
While the coherent part of the spectral function (delta function)
and its incoherent power law tail have been mixed by the convolution
into one smooth lineshape, this clearly
demonstrates both of the effects discussed above --- there is an
overall shift on the line towards lower binding energy (to the right),
and spectral weight is transfered from the
tail of the line to the peak, making it seem sharper and more symmetric.
It is this sharpening of the line,
observable provided that the charge gap (in this case
$2\Delta$) is larger than the inverse of the core hole lifetime $1/\tau$,
which is the key signature of the failure of the orthogonality catatrophe.

While this model is obviously a gross oversimplification, it
does illustrate the two main effects of opening a gap
at the fermi energy; to restate --- the delta function peak
at threshold is partially restored, and there is an overall
shift in the position of the line shape towards lower binding
energy.

\section{Superconducting Systems}
\label{sc}

\begin{figure}[tb]
\begin{center}
\leavevmode
\epsfxsize \columnwidth
\epsffile{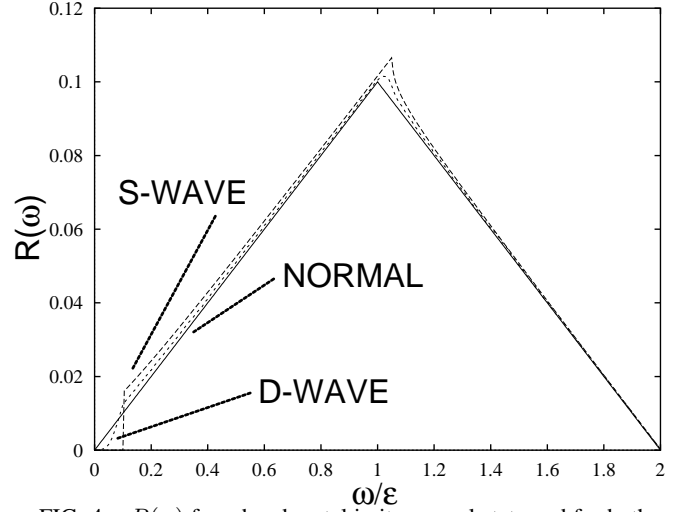}
\caption{
$R(\omega)$ for a band metal in its normal state and for both
s and d wave superconductors
with a gap of $\Delta/\epsilon=0.05$.  ($\alpha=0.1$)
}
\label{fig4}
\end{center}
\end{figure}

A similar analysis can be performed for a superconductor.
The same basic physics holds as for the toy model considered
above, but the density-density correlation function now factorizes
into normal and anomalous parts, so we must consider
\begin{eqnarray}
P(i\Omega_n) &=&
\frac{1}{\nu^2}
\sum_{p,k} |V(k,p)|^2\frac{1}{\beta} \sum_{ip_n}
[{\cal G}(p,ip_n){\cal G}(k,ip_n+i\Omega_n) \nonumber\\
&-& {\cal F}(p,ip_n){\cal F}^{\dagger}(k,ip_n+i\Omega_n) ]
\end{eqnarray}
where $i\Omega_n$ and $ip_n$ are Matsubara frequencies.

Performing the frequency sum and continuing back to real
frequencies yields the expression for $R(\omega)$ in a superconductor.
At zero temperature, and for $\omega>0$ this reduces to:
\begin{eqnarray}
R(\omega) &=& \frac{1}{\nu^2}\sum_{k,p} |V(k,p)|^2
\nonumber\\ &\times& \lbrace
(v_p^2u_k^2 + \frac{\Delta_{p}\Delta_{k}}{4E_pE_k} )\delta(\omega - E_p - E_k)
\rbrace
\end{eqnarray}
where $u_k^2=\frac{1}{2}(1 + \frac{\xi_k}{E_k})$ and
$v_k^2=\frac{1}{2}(1 - \frac{\xi_k}{E_k})$
are the coherence factors and $E_k=\sqrt{\xi_k^2 + \Delta_k^2}$
is the excitation energy.

We now examine both s and d--wave superconductors for two
different single particle energy dispersions, the linear dispersion
of the previous section and a two dimensional tight binding model,
and observe how the XPS lineshapes change upon the opening of
the superconducting gap.

\subsection{Linear Dispersion}

\begin{figure}[tb]
\begin{center}
\leavevmode
\epsfxsize \columnwidth
\epsffile{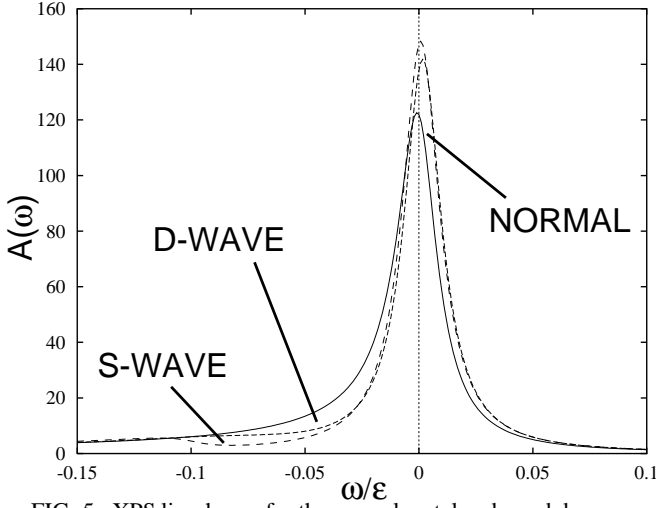}
\caption{XPS lineshapes for the normal metal and s and d wave
superconductors with a gap of $\Delta/\epsilon=0.05$.
$1/\tau = 0.01\epsilon$ and $\alpha=0.1$.  As in the band metal, the
superconductor lineshapes exhibit a delta function restoration
and a overall shift upon the opening of the gap.
}
\label{fig5}
\end{center}
\end{figure}
As before, we assume a flat single particle density of states
of the form:
\be
N(\omega) = \left\{
   \begin{array}{ll}
      N_0 & \quad \mid \omega \mid < \epsilon\\
      0   & \quad \mid \omega \mid > \epsilon\\
   \end{array} \right.
\en

\begin{figure}[tb]
\begin{center}
\leavevmode
\epsfxsize \columnwidth
\epsffile{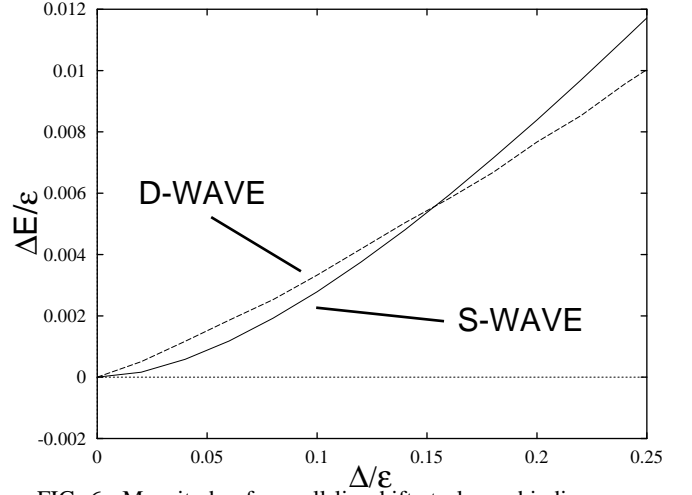}
\caption{Magnitude of overall lineshifts to lower binding energy
for s-- and d--wave superconductors assuming a linear single particle
dispersion and asymmetry exponent $\alpha=0.1$.
}
\label{fig6}
\end{center}
\end{figure}

In a BCS s--wave superconductor with a spherical
fermi surface in the normal state, the density of excitations is
given by:
\be
N(E) = \left\{ \begin{array}{ll}
N_0 \frac{E}{\sqrt{E^2 - \Delta_0^2}} &\mbox{if $E>\Delta_0$} \\
0 & \mbox{otherwise}
\end{array}
\right.
\en
Taking advantage of the delta function, an integral expression
for the $R(\omega)$ of an
s--wave superconductor is obtained.
\begin{equation}
R_S(\omega)= \alpha\int_\Delta^{\omega-\Delta}
\frac{E_p(\omega - E_p) + \Delta^2}{\sqrt{E_p^2 - \Delta^2}
\sqrt{(\omega-E_p)^2 - \Delta^2}}dE_p
\end{equation}
We plot $R_S(\omega)$ in Fig. \ref{fig4}.  The finite value of
$R_S(\omega)=\pi\Delta$ at $\omega=2\Delta$ is
a consequence of the divergant, but integrable, density of
states for $E\to\Delta$.  In the same manner as before
we can calculate the lineshape (Fig. \ref{fig5})
and shift (Fig. \ref{fig6}) for an
{\it s} wave superconductor.
The existence of the gap in a superconductor produces the same effects as
it does in the band metal. Again there is
a partial restoration of the delta function peak,
with a supression of the powerlaw tail for $\omega<2\Delta$,
and an overall shift of the line.  These effects are however
somewhat less pronounced than in the toy model.

The case of a d--wave superconductor, where gap nodes
lead to the presence of low energy excitations even for
$\Delta \ne 0$, is subtly different.  In this case the
orthogonality catastrophe fails not because of the absence
of zero energy excitations, but because the number of
zero energy particle--hole pairs produced remains countable.
More formally, the opening of a d--wave gap at zero temperature
means that the leading term in $R(\omega)$ is no longer $\alpha\omega$
but $\beta\omega^3$, and so the logarithm found from
Eqn. \ref{lineshape} in the case of a metal is eliminated.
Unlike the s--wave case, the spectral function
does have incoherent structure for $0<\omega<2\Delta$, but
this is accompanied by a slightly less pronounced transfer of spectral
weight to a delta function at threshold.

The normal state has a circular fermi surface, the superconducting
a two dimensional d-wave gap given by:
\begin{equation}
\Delta_D(\phi)=\Delta_0cos(2\phi)
\end{equation}
with $0\le\phi\le 2 \pi$.
Both $R(\omega)$ and the lineshift were obtained numerically via monte 
carlo integration.  The lineshape was calculated as above.  

The d-wave lineshape is shown along with that of the s--wave superconductor
in Fig. \ref{fig5}  and the shift in Fig. \ref{fig6}.
The presence of nodes in the gap leads to a slightly different
gap dependence of the shift in the core line; in fact
the d--wave line shift is {\it larger}
at small $\Delta$ than in the s--wave case.  The shifts can be
fitted to power laws at small $\delta = \Delta/\epsilon$. The shift for the
s--wave superconductor is given by $\Delta E_s = 1.11\alpha \delta^{1.60}$ and
the d--wave shift by $\Delta E_d = 0.51\alpha \delta^{1.18}$
with an  uncertainty of $\pm 0.01$ in both the
coefficient and the power.

If we assume that the gap opens
as $\sqrt{t}$ (where $t = (Tc - T)/Tc$),
as would be expected of a meanfield order parameter
for $T \approx T_c$, this translates into a shift in the line
scaling as $\Delta E_{s} \sim t^{0.8}$ in the s-- and
$\Delta E_{d} \sim t^{0.25}$ in the d--wave case.

\subsection{Tight Binding Model}

In order to make closer contact with real HTc superconductors,
we also evaluated lineshapes and shifts for a d--wave
superconductor on a half--filled square lattice with
underlying tight binding electron dispersion
\be
\epsilon(k_x,k_y) = -2t(cosk_x + cosk_y) -4t'cosk_x cosk_y
\en
where $t$ and is the nearest neighbour and $t^{\prime}$ the
next nearest neighbour hopping integral.  We chose 
$t'/t=-0.35$ as being
representative of the in--plane $Cu$ $d$ band in a 'standard'
123 compound such as $YBa_2Cu_3O_{7-\delta}$.  A 'standard'
2212 compound such as $Bi_2Sr_2CaCu_2O_{8+x}$ would have
$t'/t=-0.2$ \cite{bobnben} but this will not change our results
significantly.

\begin{figure}[tb]
\begin{center}
\leavevmode
\epsfxsize \columnwidth
\epsffile{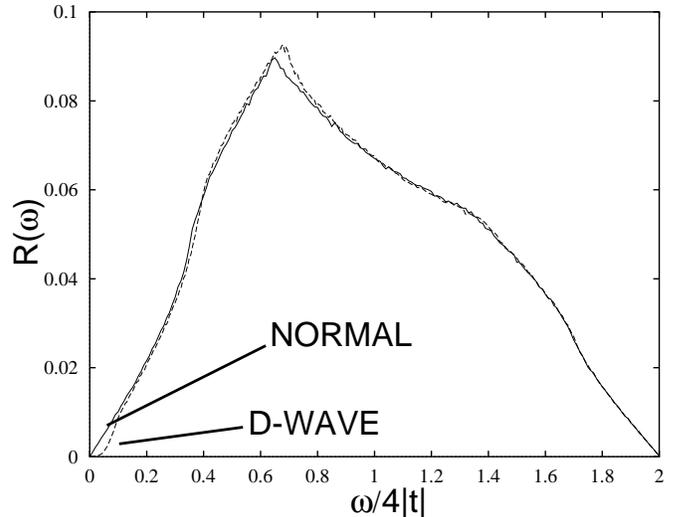}
\caption{$R(\omega)$ for tight binding metal with
$t^{\prime}/t=-0.35$ in the normal
state and in a d--wave superconducting state
with $\Delta_0/4|t| = 0.05$.  Once again, $\alpha=0.1$.
}
\label{fig7}
\end{center}
\end{figure}

We model a superconductor with d--wave symmetry on a square
lattice with a gap function of the form
\be
\Delta_{d}(k_x,k_y) &=& \frac{\Delta_0}{2}(cos(k_x)-cos(k_y))
\en
The results for $R(\omega)$ for this model of the normal and
superconducting state are shown in Fig. \ref{fig7}.
Once again we have set the coefficient of the leading linear term
in the tight binding metal to be $0.1$, so that the XPS
asymmetry exponent for the system without gap is $\alpha=0.1$.
For comparison with our previous models we consider a gap size 
of $\Delta_0 = 0.05(4|t|)$.  In general, $t$ is of the order
of $0.25 meV$. This gives $\Delta_0=50 meV$
which is clearly larger than in the real compounds, but
not unreasonably so; estimates for 
YBCO yield $\Delta_0 \approx 16 meV$ and for BSSCO 
$\Delta_0 \approx 30 meV$ \cite{shen}.

\begin{figure}[tb]
\begin{center}
\leavevmode
\epsfxsize \columnwidth
\epsffile{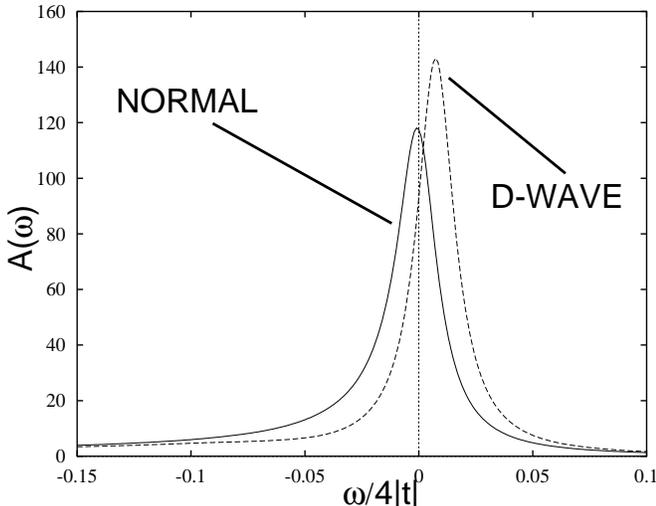}
\caption{XPS lineshapes for a tight binding metal in the normal
state and in a
d--wave superconducting state.   $\Delta_0/4|t| =0.05$, 
$1/\tau=0.01|4t|$
and $\alpha=0.1$.
}
\label{fig8}
\end{center}
\end{figure}

\begin{figure}[tb]
\begin{center}
\leavevmode
\epsfxsize \columnwidth
\epsffile{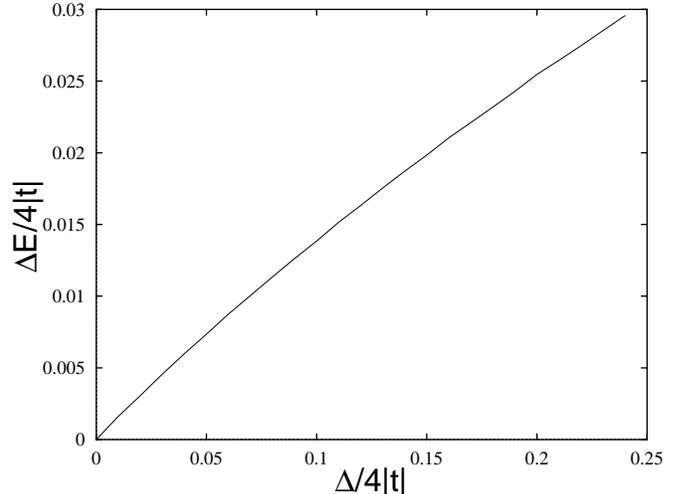}
\caption{Overall line shifts to lower binding energy
for a d--wave superconductor with
a tight binding dispersion in the normal state as a function
of the gap parameter $\Delta_0/4|t|$, once again $\alpha=0.1$.
}
\label{fig9}
\end{center}
\end{figure}

The resulting lineshapes and shifts for these coefficients are
displayed in Figs. \ref{fig8} and \ref{fig9}.   Clearly
the overall trends are exactly the same as found in the
more general models; a sharpening of the
XPS line and a shift to lower binding energy, but both trends
are somewhat more marked than in the constant density of states
model with the same parameters
(\cf Figs. \ref{fig5} and \ref{fig6}),
and the shift is now very nearly linear in $\Delta_{0}$.

The size of shift which we find must be something of an underestimate
of the true shift in the cuprates, since we neglect all corrections
to screening of the core hole which arise from electron--electron
interaction, which is known to be strong in these systems.  An improved
estimate could be found by incorporating a Hubbard $U$ term in the
model, and evaluating an RPA series for the screened susceptibility.
Such a proceedure has been found to be necessary to obtain
quantative estimates of the local spin susceptility in
these systems \cite{bob}.
Screening through electron--electron interaction is of course a
dynamical process, and
leads to corrections to the lineshape and gap dependance of the
shift, as well as to its overall scale.  These effects do {\it not}
change the underlying physics, but
can in principle be included in our calculation scheme, and may need to
be included in any serious quantitive attempt to fit experimental
lineshapes and shifts.
We note also that attempts to fit the modified asymmetric
lineshape found which we find in the presence of a gap with a
standard Doniach--Sunjic lineshape can lead to artificially low
values of $\alpha$.

\section{Pseudogaps and comparison with experiment}
\label{pseudogaps}

Pseudogap behaviour, which can be loosely defined as the
(partial) loss of low energy excitations without the emergence of
order, has been observed in many strongly correlated
electron systems \cite{footnote2}.  The best known example is provided  by
the underdoped cuprate superconductors, where evidence
for the opening of a pseudogap is found from NMR and ARPES
experiments at temperatures between some high energy scale
$T^*$ and the superconducting transition temperature $T_c$.
As a function of doping, $T^*$ interpolates between the
N\'eel temperature $T_N \sim 800K$ of the undoped Mott
insulator and the transition temperature $T_c \sim 100K$ of the
optimally doped superconductor.

What could XPS teach us about the opening of a pseudogap in
this case ?  If the pseudogap seen in NMR is a precursor
to the formation of superconducting order at low
temperatures, coming about, for example, through the
formation of ``incoherent'' pairs of electrons, then its
effects on XPS lines will be broadly the same as those
for a true gap.  If on the other hand the pseudogap
is not a precursor to superconductivity but, as has been
suggested, a gap to spin excitations only, then XPS
spectra would undergo little or no change at $T^*$.
In this way the characteristic XPS signatures of gap
opening --- the sharpening of an asymmetric core line
and/or a shift of lines to lower binding energy ---
could be used to distinguish between different theories
of pseudogapped systems.

A shift in core lines to lower binding energy
{\it has} been reported for XPS spectra taken above and below
$T_c$, {\it and} for spectra taken above and below $T^*$
in the cuprate high temperature
superconductor $Bi_2Sr_2Ca_{1-x}Y_xCu_2O_{8+\delta}$ \cite{karlson}.
At first sight this seems to offer confirmation of exactly the
type of effect which we predict on the basis of core hole
screening.
However, the quoted experimental values of the shift are of order
$100 meV$ for lines with an asymmetry $\alpha = 0.04$.  
This is much larger
than can be reconciled with our simple model; using the tight binding
model considered above, together with the parametrization
$\alpha = 0.04$ and $\Delta_0 = 30 meV$, we would anticpate a 
shift in the XPS line of order $1.2 meV$.

Of course, analyzing the metal superconductor
transition in $Bi_2Sr_2Ca_{1-x}Y_xCu_2O_{8+\delta}$ in terms
of tight binding model and BCS models will not
always give reliable answers, especially in the underdoped
``pseudogap'' regime.
Nevertheless we believe our calculation provides the correct
starting point for understanding such experiments, and mechanisms 
other than the supression of screening of the core hole by the opening
of the gap should therefore also be considered.

\section{Conclusions}
\label{conclusions}

The gross effects seen in XPS lineshapes for metallic systems
when a gap opens are very robust and independent of the choice
of model.  The core line undergoes an overall shift to lower
binding energy and spectral weight is transfered from the powerlaw
tail of the line to its peak, leading to a sharpening of the line
and some loss of overall asymmetry.
Modifications to lineshape are more easily seen in systems
where the gap is large compared with the intrinsic width of
the core level.  In this limit subtle differences can be also seen
between different models.

The shift in core lines shows power law dependence on the
size of the gap, with different power laws for superconductors with
s-- and d--wave
gap symmetries and for a tight binding model as compared with
a model with a constant density of states.  In principle
XPS offers a means of distinguishing between different
gap symmetries in systems with complicated order parameters.

Where sufficiently narrow core lines can be found, XPS
offers a potentially rich source of information about
the changes which take place in many electron systems
when a gap opens, and might be particularly useful if taken
in parallel with NMR measurements.  The opening of
pseudogaps could also be studied from the perspective of XPS
measurements.

For simplicity, we have chosen to work within perturbation
theory and to discuss only models which have simple
non--interacting quasi--particle excitations.  Both
of these restrictions can be relaxed, and
many of the same physical considerations apply to
core levels coupled to strongly interacting electron systems.
Experimentally it might well be interesting to look
at the effect on XPS lines of metal--insulator or charge
density wave transitions where the intrinsic gap scale
is very much larger, and the effects of gap opening can
be expected to be more pronounced.

{\it Acknowledgments.}
It is our pleasure to acknowledge many helpful converstaion with
Jim Allen, Franz Himpsel, and Robert Joynt. 
This work was supported under the grant DMR--9704972 .

\end{document}